\documentclass[conference]{IEEEtran}
\IEEEoverridecommandlockouts
\usepackage{cite}
\usepackage{bbold}
\usepackage{amsmath,amsthm, mathtools, amssymb, amsfonts, amsbsy,fixmath}
\usepackage[capitalise]{cleveref}
\usepackage{algorithmic}
\usepackage{graphicx}
\usepackage{textcomp}
\usepackage{xcolor}
\usepackage[ruled,vlined]{algorithm2e}
\def\BibTeX{{\rm B\kern-.05em{\sc i\kern-.025em b}\kern-.08em
    T\kern-.1667em\lower.7ex\hbox{E}\kern-.125emX}}

\def\cbr#1{\left\lbrace #1 \right\rbrace} 
\def\sbr#1{\biggl [ #1 \biggr ]} 
\def\nbr#1{\left( #1\right)}
\def\bs #1{\boldsymbol{#1}}
\def\bm#1{\mathbf{#1}}
\newcommand{\cB}[1]{\textcolor{black}{#1}}

    \def\p {{(p)}}
\def\n {{n'}}
\begin{document}

\title{Online Non-linear Topology Identification from Graph-connected Time Series\\
\thanks{This work was supported by the IKTPLUSS INDURB grant 270730/O70 , SFI Offshore Mechatronics grant 237896/O30 and the PETROMAKS Smart-Rig grant 244205/E30 from the Research Council of Norway.}}
% Kernal-based Online Non-linear Topology Identification for VAR Graph-connected Time series

\author{
\IEEEauthorblockA{\textit{Rohan Money, Joshin Krishnan, Baltasar Beferull-Lozano}\\
\textit{WISENET Center, Department of Information and Communication Technologies} \\
\textit{University of Agder,}
Grimstad, Norway \\
\{rohan.t.money; joshin.krishnan; baltasar.beferull\}@uia.no}}

\maketitle

\begin{abstract}
 Estimating the unknown causal dependencies among graph-connected time series plays an important role in many applications, such as sensor network analysis, signal processing over cyber-physical systems, and finance engineering.  Inference of such causal dependencies, often know as topology identification, is not well studied for non-linear non-stationary systems, and most of the existing methods are batch-based which  are not capable of handling streaming sensor signals. In this paper, we propose an online kernel-based  algorithm for topology estimation of non-linear vector autoregressive time series by solving a sparse online optimization framework using the composite objective mirror descent method. Experiments conducted on real and synthetic data sets show that the proposed algorithm outperforms the state-of-the-art methods for topology estimation. 
\end{abstract}


\section{Introduction}
\cB{Recent advancements in cyber-physical systems (CPS) and sensor networks call for advanced research on data analysis of structured or inter-linked spatio-temporal signals. Such structured signals can be meaningfully represented using graph-connected time series. Graph representation is a prevalent tool to model the inter-dependency of data \cite{leax.d2019}, and it plays a vital role in countless practical applications such as time  series prediction \cite{onlB.Z2017}, change point detection \cite{dynl.m2018}, data compression \cite{pata.c2018}, etc.}

\cB{Many of the functional dependencies in real-world time series are causal \cite{equc.j2011}, and inferring the causal dependencies, which we term as \textit{topology identification}, generates a more informative representation of the multivariate  data. These dependencies may not be physically observable in some cases; instead, there can be logic connections between data nodes that are not physically connected due to control mechanisms, and inferring such typologies is a challenging task. Linear models, such as structural equation models (SEM), vector auto-regressive (VAR) models, and structural vector auto-regressive (SVAR) models \cite{topg.b2018} are widely used to study the causal dependencies among the graph-connected time series. SEM being a memory-less model, does not accommodate the temporal dependencies among the data, whereas the VAR is an ideal choice for modeling the time-lagged interactions; however, it fails to capture the instantaneous causal relations. SVAR is a slightly modified model that unifies both SEM and VAR. The choice of the model depends on the physical nature of the system; for instance, SVAR is a useful model for brain connectivity analyses. However, VAR deserves special attention since the nodal dependencies on many practical sensor networks (e.g., water networks, oil and gas networks)  involve mainly time-lagged interactions.}

\cB{A significant challenge connected to topology identification is that the real-world systems are usually non-stationary, meaning that the statistical properties of dependencies vary over time. The commonly used batch-based off-line methods \cite{nonshe2019} have two major drawbacks: \textbf{i)} they are not effective in tracking the topology of non-stationary systems and \textbf{ii)} from a pure computational point of view, they suffer from processing large batch of data; hence, it is necessary to develop online estimation algorithms \cite{onlzam2019}.  Online topology estimation algorithms have been developed for linear models, meaning that the causal dependencies among the data  time-series hold a linear relation. For instance, in  \cite{onlzam2019}, a novel online linear topology identification algorithm have been proposed by minimizing a group-lasso-regularized \cite{aspnoa2013} objective function.}

\cB{Although the linear topology identification is a well-studied problem, many practical systems have non-linear dependencies \cite{vecche2011}. As an example, in a smart water network, the causal dependencies are non-linear due to various control systems, saturation in valves or pumps, and non-linear physical equations  governing the system.  Similarly,  essential non-linear dependencies are present in most of the real-world systems such as brain networks and finance networks. The ability of nonparametric techniques \cite{unimic2006} and deep neural networks to learn  non-linear functions is well studied, which has been exploited also in topology identification  \cite{neuale2018}, \cite{nonshe2019}, \cite{kery.s2018}. However, once again most of these algorithms are batch-based.}

\cB{Kernel-based representations are powerful tools to model the non-linear dependencies \cite{leasch2001}, which can be exploited to develop algorithms for online non-linear topology identification. For instance, in \cite{only.s2018}, authors have proposed an online algorithm based on functional gradient descent by considering a SVAR model. In \cite{onlm.m2020}, authors used a more general non-additive model for topology identification and a dictionary-based approach to solve the computational complexity imposed by the kernels. But \cite{onlm.m2020} restricts the choice of the kernel functions to be twice differentiable to learn a sparse topology.}

\cB{This paper proposes an online topology identification algorithm based on a non-linear VAR model using kernels. The proposed algorithm learns sparse and time-varying non-linear typology by solving an online optimization framework using composite objective iterations \cite{Comjoh2010}. We provide strong empirical evidence using real and synthetic data sets, which show that the proposed algorithm outperforms its state-of-the-art counterparts.}

\section{Problem Formulation}
%\subsection{Non-linear topology identification}
\cB{Consider a collection of $N$ time series, connected by a directed graph and let $y_n[t]$ be the value of time series at time $t=0,1,\dots,T-1$ measured at node $1\leq n\leq N$.}
A $P$-th order non-linear VAR model of the time series can be formulated as
	\begin{align} \label{eq:var}
	y_n[t] =   \sum_{\n = 1}^{N} \sum_{p = 1}^{P}a_{n,\n}^{\p} f_{n,n'}^{(p)}(y_{\n}[t-p])+u_n[t],
	\end{align}
where $f_{n,n'}^{(p)}$ is a non-linear function that captures the causal influence of the $p$-lagged data at node $\n$ on the node $n$, \cB{$a_{n,\n}^{(p)}$ is the corresponding entry of the graph adjacency matrix, and $u_n[t]$ is the measurement noise.} Referring to \eqref{eq:var}, topology identification can be defined as the estimation of \cB{the non-linear dependencies expressed by $\cbr{a_{n,\n}^{\p}f_{n,n'}^{(p)}(.)}_{p=1}^P$} for $n=1,2,\dots,N$ from the observed time series $\cbr{y_n[t]}_{n=1}^N$.

\cB{To circumvent the challenges in topology identification, imposed by the non-linear dependencies, we assume that the functions $f_{n,n'}^{(p)}(.)$ in \eqref{eq:var} belong to a reproducing kernel Hilbert space (RKHS):}

 \resizebox{1\linewidth}{!}{
	\begin{minipage}{1.09\linewidth}
\begin{align}\label{eq:rkhs}
\mathcal{H}_{n'}^\p:=\cbr{{f}_{n,n'}^{\p}|{f}_{n,n'}^{\p}\nbr{y}=\sum_{t=0}^{\infty}\beta^\p_{n,n',t}~\kappa^\p_{n'}\nbr{y,y_\n[t-p]}},
\end{align}
\end{minipage}
}
 \cB{where $\kappa^\p_\n: \mathbb{R}\times \mathbb{R} \rightarrow \mathbb{R}$ is the Hilbert space basis function, often known as the kernel, which measures the similarities between the arguments of the basis function. Using \eqref{eq:rkhs}, a function ${f}_{n,n'}^{\p}$ evaluated at $y$ can be represented as the linear weighted sum of the similarities between $y$ and the data samples $\cbr{y_\n[t-p]}_{t=0}^{t=\infty}$, where the weights are denoted by $\beta^\p_{n,n',t}$. We assume that the Hilbert space is characterized by the inner product $\langle \kappa^\p_\n(y,x_1),\kappa^\p_\n(y,x_2)\rangle:=\sum_{t=0}^{\infty}\kappa^\p_\n(y[t],x_1)\kappa^\p_\n(y[t],x_2)$, with the kernel having the reproducible property $\langle\kappa^\p_\n(y,x_1),\kappa^\p_\n(y,x_2)\rangle=\kappa^\p_\n(x_1,x_2)$. Such a Hilbert space with the reproducing kernel constitutes an RKHS with norm $\|{f}_{n,n'}^{\p} \|^2_{\mathcal{H}_{n'}^\p} =\sum_{t=0}^{\infty}\sum_{t'=0}^{\infty}\beta^\p_{n,n',t}~\beta^\p_{n,n',t'}~\kappa^\p_\n(y_n[t],y_n[t'])$.  We refer to \cite{Wahba1990} for further reading on RKHS.}
 
For a node $n$, the least-squares (LS) estimates of $\cbr{{f}_{n,\n}^{\p}\in \mathcal{H}_{n'}^\p;\n=1,\dots,N,~p=1,\dots,P}$ are obtained by solving the following non-parametric optimization problem:
 \begin{align} \label{eq: ls}
\cbr{\widehat{{f}}_{n,\n}^\p}_{\n,p}=
\arg &\min_{\cbr{{f}_{n,\n}^{\p}\in \mathcal{H}_\n^\p}} \frac{1}{2}\sum_{\tau=P}^{T-1}\sbr{y_{n}[\tau]- \nonumber\\ &\sum_{\n = 1}^{N} \sum_{p = 1}^{P}a_{n,\n}^\p {f}_{n,\n}^\p(y_\n[\tau-p])}^2.
\end{align}
\cB{It is to be noted that, in \eqref{eq: ls}, the functions $\{{f}_{n,\n}^{\p}\}$ belongs to the RKHS, defined in \eqref{eq:rkhs}, which is an infinite dimensional space. 
However, by resorting to the Representer Theorem} \cite{ageolk2000}, \cB{the solution of \eqref{eq: ls} can be written using a finite number of data samples:}  
\begin{align}\label{eqn:rep}
    \widehat{f}_{n,n'}^{\p}\nbr{y_\n[\tau-p]}~~~~&\nonumber\\
    =\sum_{t=p}^{p+T-1}\beta^\p_{n,n',(t-p)}&\kappa^\p_{n'}\nbr{y_\n[\tau-p]),y_\n[t-p]}.
\end{align}
\cB{Using \eqref{eqn:rep}, \eqref{eq: ls} can be reformulated as a parametric optimization problem involving the available data samples, as follows: 
 \begin{align} \label{eq: ParOpt}
\cbr{\widehat{\alpha}_{n,\n,t}^\p}_{\n,p,t}=
\arg &\min_{\cbr{\alpha_{n,\n,t}^\p}} \mathcal{L}^n\nbr{\alpha_{n,\n,t}^\p},
\end{align}
where}
\begin{align} \label{eq: ls2}
\mathcal{L}^n\nbr{\alpha_{n,\n,t}^\p}&:= \frac{1}{2}\sum_{\tau=P}^{T-1}\sbr{y_{n}[\tau]- \sum_{\n = 1}^{N} \sum_{p = 1}^{P}\sum_{t=p}^{p+T-1}{\alpha}^\p_{n,n',t}\kappa^\p_{n'}\nbr{\tau,t}}^2,\\
    \alpha_{n,\n,t}^\p&:=a_{n,\n}^\p\beta_{n,\n,(t-p)}^\p,  \label{eqn:alpha}\\
    \text{and}~~~~~~~~~&\nonumber\\
    \kappa^\p_{n'}\nbr{\tau,t}&:=\kappa^\p_{n'}\nbr{y_\n[\tau-p]),y_\n[t-p]}. \label{eqn:kappa}
\end{align}

\cB{We stack the entries of $\cbr{{\alpha}^\p_{n,n',t}}$ and $\cbr{\kappa^\p_{n'}\nbr{\tau,t}}$ in the lexicographic order of the indices $p$, $\n$, and $t$ to obtain the vectors $\bs {\alpha}_n\in \mathbb{R}^{PNT}$ and $\boldsymbol{\kappa}_\tau\in \mathbb{R}^{PNT}$, respectively, and rewrite \eqref{eq: ParOpt} as
 \begin{align} \label{eq: ParOpt2}
\widehat{\bs{\alpha}}_n=
\arg &\min_{\bs\alpha_{n}}  \mathcal{L}^n\nbr{\bs{\alpha}_{n}},
\end{align}
where
 \begin{align} \label{eq: ls4}
\mathcal{L}^n(\bs{\alpha}_n)=\frac{1}{2}\sum_{\tau=P}^{T-1}\sbr{y_{n}[\tau]- \bs{\alpha}_{n}^\top\bs{\kappa}_\tau}^2
\end{align}
Further, to avoid overfitting and to enforce group sparsity, we propose a regularized optimization framework:
 \begin{align} \label{eq: ParOpt2}
\widehat{\bs{\alpha}}_n=
\arg &\min_{\bs\alpha_{n}}  \mathcal{L}^n\nbr{\bs{\alpha}_{n}}+\lambda \sum_{\n = 1}^{N} \sum_{p = 1}^{P}\| \bs{\alpha}_{n,n'}^\p \|_2,
\end{align}
where $\lambda\geq0$ is the regularization parameter and $\bs{\alpha}_{n,n'}^\p =(\alpha_{n,n',0}^\p,\alpha_{n,n',1}^\p,\dots,\alpha_{n,n',T}^\p)\in \mathbb{R}^T.$
The second term in \eqref{eq: ParOpt2} is a \textit{group-lasso} regularizer, which promote a \textit{group-sparse structure} in $\bs{\alpha}_{n,n'}^\p$, thereby exploiting the prior information that the number of causal dependencies are typically small for  real-world graph-connected time series.}

\cB{The parametric optimization given by \eqref{eq: ParOpt2} is a batch (offline) solver meaning that to solve \eqref{eq: ParOpt2}, we require all data samples $\cbr{y_n[\tau]}_{\tau=P}^{T-1}$ to be available. Such an offline approach has two major drawbacks: \textbf{i)} it is not suitable for real-time applications since the solver has to wait for the entire batch of data and \textbf{ii)} it suffers from high computation complexity and memory requirements which grows super linearly with the batch size. In the following section, we propose an online algorithm to estimate the coefficients $\bs{\alpha}_{n}$ in \eqref{eq: ParOpt2}.}
\section{Online topology estimation}
    \cB{First replace the original loss function $\mathcal{L}^n(\bs\alpha_n)$ in \eqref{eq: ParOpt2} with the instantaneous loss function $l_\tau^n(\bs\alpha_n)=\frac{1}{2}[y_{n}[\tau]- \bs{\alpha}_{n}^\top\bs{\kappa}_\tau]^2$:
     \begin{align} \label{eq: ParOpt3}
\widehat{\bs{\alpha}}_n=
\arg &\min_{\bs\alpha_{n}}  l_\tau^n\nbr{\bs{\alpha}_{n}}+\lambda \sum_{\n = 1}^{N} \sum_{p = 1}^{P}\| \bs{\alpha}_{n,n'}^\p \|_2.
\end{align}
A straightforward way to solve \eqref{eq: ParOpt3} is by applying the online subgradient descent (OSGD). However, it is to be remarked that the regularizer in \eqref{eq: ParOpt3} is non-differentiable and OSGD fails to provide sparse $\bs{\alpha}_{n,n'}^\p$ since it linearizes the entire instantaneous objective function in \eqref{eq: ParOpt3} \cite{onlzam2019}.} To mitigate this issue, we use the composite objective mirror descent (COMID) \cite{Comjoh2010} algorithm. The online COMID update can be written as 
    \begin{align}\label{eqn:comid_update}
    {\bs{\alpha}}_n[t+1]&=\arg \min_{{\bs{\alpha} }_n} J_t^{(n)}({\bs{\alpha}}_n),
    \end{align}
 where
    \begin{align}
    J_t^{(n)}({\bs{\alpha}}_n) &\triangleq \nabla \ell_{t}^n({\bs{\tilde{\alpha}}}_n[t])^\top\nbr{{\bs{\alpha}}_n -{\bs{\tilde{\alpha}}}_n[t]}\nonumber\\ &+\frac{1}{2\gamma_t}\| {\bs{\alpha}}_n -{\bs{\tilde{\alpha}}}_n[t]\|_2^2+\lambda \sum_{\n = 1}^{N} \sum_{p = 1}^{P}\| \bs{\alpha}_{n,n'}^\p \|_2. \label{eqn:comid_risk}
\end{align}
In \eqref{eqn:comid_risk}, ${\bs {\tilde{\alpha}}}_n[t]\in \mathbb{R}^{PN(t+1)}$ is defined as $[\bs \alpha_n[t];\bm{0}]$, where $\bs \alpha_n[t]\in \mathbb{R}^{PNt}$ is the value of $\bs \alpha_n$  estimated by processing the samples up to time $t$. The zero vector $\mathbf{0}\in \mathbb{R}^{PN}$ is appended as an initialization for the coefficients of the new elements of the kernel vector corresponding to the $(t+1)^{th}$ data sample. In \eqref{eqn:comid_risk}, the first term is the gradient of the loss function and the second term is the Bregman divergence $B({\bs{\alpha}}_n ,{\bs{\tilde{\alpha}}}_n[t])=\frac{1}{2}\| {\bs{\alpha}}_n -{\bs{\tilde{\alpha}}}_n[t]\|_2^2$, chosen in such a way that the COMID update has a closed form solution \cite{bregut2011} and $\gamma_t$ is the corresponding step size. Bregman divergence ensures that ${\bs \alpha}_n[t+1]$ is close to ${\bs{ \tilde{\alpha}}}_n[t]$, in line with the assumption that the topology changes smoothly. The third term is a sparsity enforcing regularizer, in order to promote sparsity in the updates. The gradient in \eqref{eqn:comid_risk} is evaluated as
\begin{align}
{\bf v}_n[t]&:=\nabla \ell_{t}^n({\bs{\tilde{\alpha}}}_n[t])= \boldsymbol{\kappa}_\tau\nbr{\bs \alpha_{n}^\top\bs\kappa_\tau-y_{n}[\tau]}
% {\bf v}_n[t]&={\bf \boldsymbol{\kappa}}_\tau\nbr{\bs \alpha_{n}^\top\bs\kappa_\tau-y_{n}[\tau]}
\label{eqn:grad}
\end{align}
 Expanding the objective function in \eqref{eqn:comid_risk} by  omitting the constants leads to the following formulation:

 \resizebox{1\linewidth}{!}{
	\begin{minipage}{\linewidth}
\begin{align} 
    J_t^{(n)}({\bs{\alpha}}_n) &\propto \frac{{\bs{\alpha}}_n^\top{\bs{\alpha}}_n}{2\gamma_t}+{\bs{\alpha}}_n^\top\nbr{{\bf v}_n[t]-\frac{1}{\gamma_t}{\bs{ \tilde{\alpha}}}_n[t]} +\lambda \sum_{\n = 1}^{N} \sum_{p = 1}^{P}\| \bs{\alpha}_{n,n'}^{(p)} \|_2\nonumber
    \end{align}
\end{minipage}
}

\begin{align}
      =\sum_{\n=1}^N \sum_{p = 1}^{P}&\sbr{\frac{{{\bs \alpha}_{n,\n}^{\p}}^\top {\bs \alpha}_{n,\n}^{\p}}{2\gamma_t}+{{\bs \alpha}_{n,\n}^{\p}}^\top\nbr{{\bf v}_{n,\n}^{\p}[t]-\frac{1}{\gamma_t}{\bs {\tilde{\alpha}}}_{n,\n}^{\p}[t]}~~~~~\nonumber\\
      &+\lambda  \| \bs \alpha_{n,n'}^{(p)} \|_2}.\label{eqn:comid_risk3}
\end{align}
Note that \eqref{eqn:comid_risk3} is separable in $\n$, $m$ and $p$. Using \eqref{eqn:comid_risk3}, a closed form solution of \eqref{eqn:comid_update} can be obtained in terms of multidimensional shrinkage-thresholding operator \cite{amupui2009} as
\begin{align}
    {\bs \alpha}_{n,\n}^{(p)}[t+1]&=\nbr{{\bs {\tilde{\alpha}}}_{n,\n}^{(p)}[t]-\gamma_t{\bf v}_{n,\n}^{(p)}[t]}\times \nonumber 
    \\ &\sbr{1-\frac{\gamma_t\lambda~\mathbb{1}\cbr{n\neq\n}}{\|{\bs {\tilde{\alpha}}}_{n,\n}^{(p)}[t]-\gamma_t{\bf v}_{n,\n}^{(p)}[t]\|_2}}_+,
    \label{eqn:sol}
\end{align}

where $[x]_+=\max\cbr{0,x}$ and 
\[
    \mathbb{1}\cbr{n\neq\n}= 
\begin{cases}
    1,& \text{if } n\neq\n\\
    0,              & n=\n.
\end{cases}
\]

The term ${\bs {\tilde{\alpha}}}_{n,\n}^{(p)}[t]-\gamma_t{\bf v}_{n,\n}^{(p)}[t]$ in \eqref{eqn:sol} performs a stochastic gradient update of $\bs {\alpha}_{n,\n}^{(p)}$ in a direction that decreases the instantaneous loss function $l_\tau^n(\bs\alpha_n)$ and the second term in \eqref{eqn:sol} promotes group sparsity of $\bs {\alpha}_{n,\n}^{(p)}$. The function $\mathbb{1}\cbr{n\neq\n}$ in the second term prevents the enforcement of sparsity of self-connections of the graph. One major issue with \eqref{eqn:sol} is that the size of $\bs{v}_{n,\n}^{(p)}[t]$ becomes prohibitive as $t$ increases. To mitigate this issue we select the recent $T_w$ data points to calculate \eqref{eqn:sol}. For the experiments presented in this paper, we heuristically fix the value of $T_w$ to 2000. Although this sub-optimal approach affects the performance of the algorithm, we are getting quite competitive empirical performance as shown later in the experiment section. 

The proposed algorithm, termed as \textit{Nonlinear Topology Identification via Sparse Online learning} (NL-TISO), is summarized in Algorithm \ref{alg:NL-TISO}.
\begin{algorithm}[h!]
 \label{alg:NL-TISO}
\SetAlgoLined
\KwResult{$\boldsymbol{\alpha}_{n,n'}^\p,  for~ n,n'=1,..,N$ and~$p=1,..,P$ }
 \textbf{Store} $\{\boldsymbol{y}_n[t]\}_{t=1}^P$,\\
 \textbf{Initialize} $\lambda$, $\gamma$ (heuristically chosen) and  kernel parameters depending on the type of the kernel. \\
 \For{$t=P,P+1,\dots$}{
  Get data samples ${y}_n[t],~\forall n$ and compute $\boldsymbol{\kappa}_\tau$
  \\\For{$n=1,\dots,N$}{
  compute $\bm{v}_n[t]$ using \eqref{eqn:grad}
 \\ \For{$\n=1,\dots, N$}{
  compute $\boldsymbol{\alpha}_{n,n'}^\p[t+1] $ using \eqref{eqn:sol} }
  }
 }
 \caption{NL-TISO Algorithm}
\end{algorithm}

\section{Experiments}
\cB{In this section, we illustrate the effectiveness of the proposed NL-TISO algorithm using synthetic and real data. We compare our results with two state-of-the-art topology estimation algorithms: \textbf{i)} TIRSO \cite{onlzam2019}- a recent online topology estimation algorithm based on COMID update developed for linear causal dependencies and \textbf{ii)} functional gradient descent (FGD) algorithm \cite{only.s2018}- an online kernal based topology estimation algorithm based on functional gradient descent updates} 
\subsection{Experiments using synthetic data}
\subsubsection{Identifying causal dependencies }
\label{Tid}
\cB{We generated graph connected time series, based on the non-linear VAR model \eqref{eq:var} with parameter values $N=5$, $T=3000$, and $P=2$. The entries of the graph adjacency matrix $\cbr{a_{n,\n}^\p}$ are drawn from a Gaussian distribution $\mathcal{N}(8,\,3)$ with an edge probability $p_e=0.1$. The initial $P$ samples of the time series are drawn randomly from a Gaussian distribution $\mathcal{N}(0,\,0.1)$ and the remaining samples are generated using model \eqref{eq:var}. A Gaussian kernel centered at the dependent data points and having variance 0.03 is used to model the non-linear dependencies in \eqref{eq:var}, where the kernel coefficients $\beta_{n,n'}^{(p)}$ are drawn from a zero mean Gaussian distribution with variance $0.03$. The noise $u_n[t]$ is generated from a zero mean Gaussian distribution with variance $0.01$. }
\begin{figure}[hbt!]
\centerline{\includegraphics[scale=0.165]{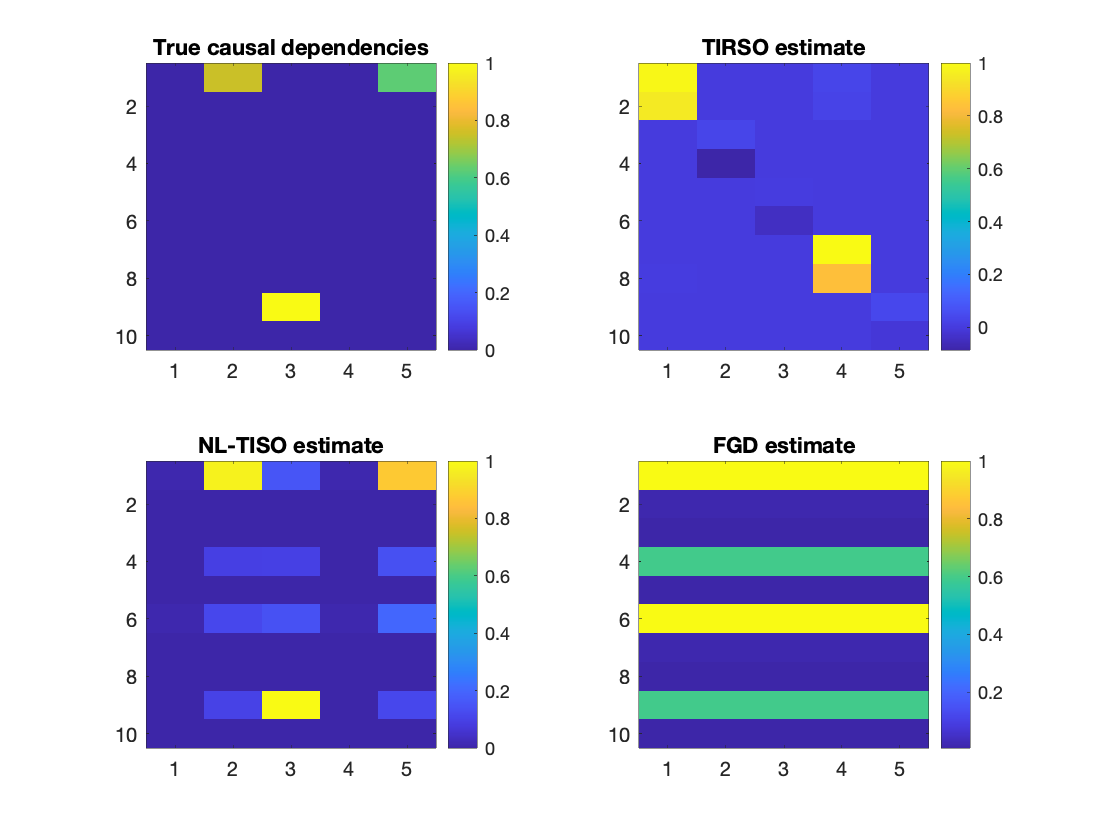}}
\caption{Causal dependencies (normalized) estimated using different algorithms compared with the  true dependency.}
\label{topid}
\end{figure}
\begin{figure}[hbt!]
\centerline{\includegraphics[scale=0.165]{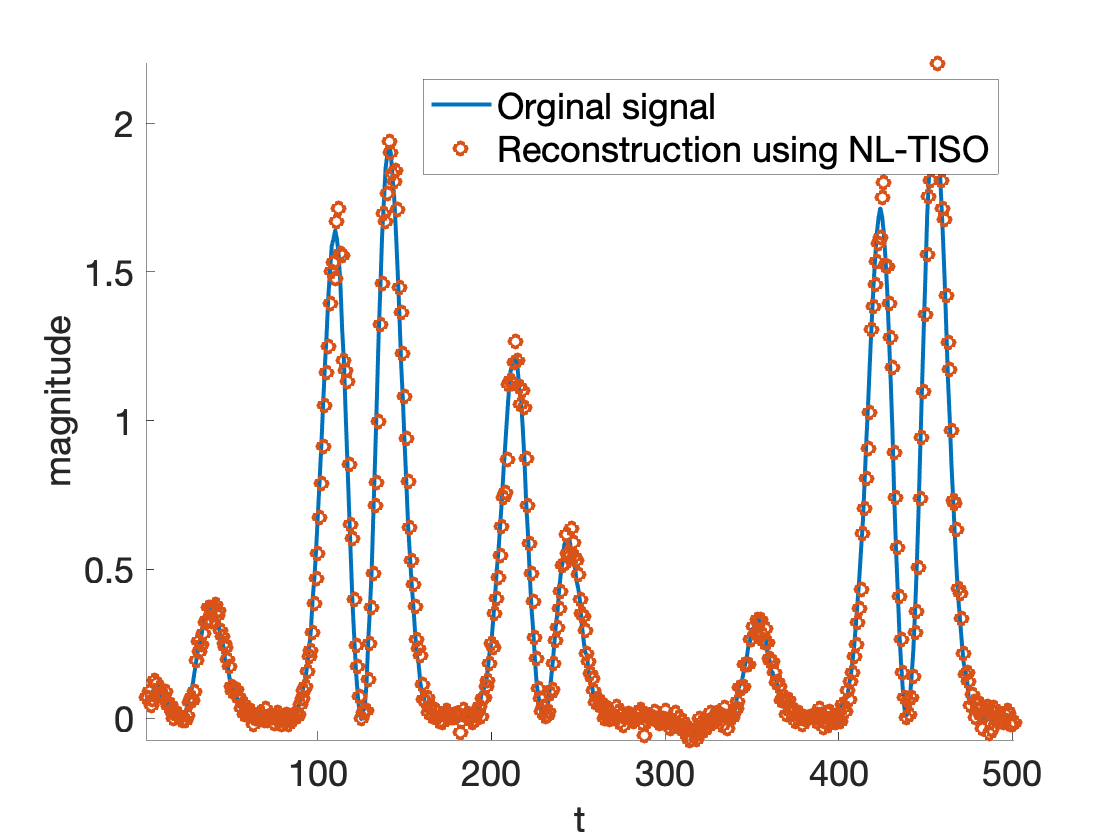}}
\caption{Reconstruction of true signal in node 1 using estimated coefficients. }
\label{rec}
\end{figure}

\begin{figure}[hbt!]
\centerline{\includegraphics[scale=0.165]{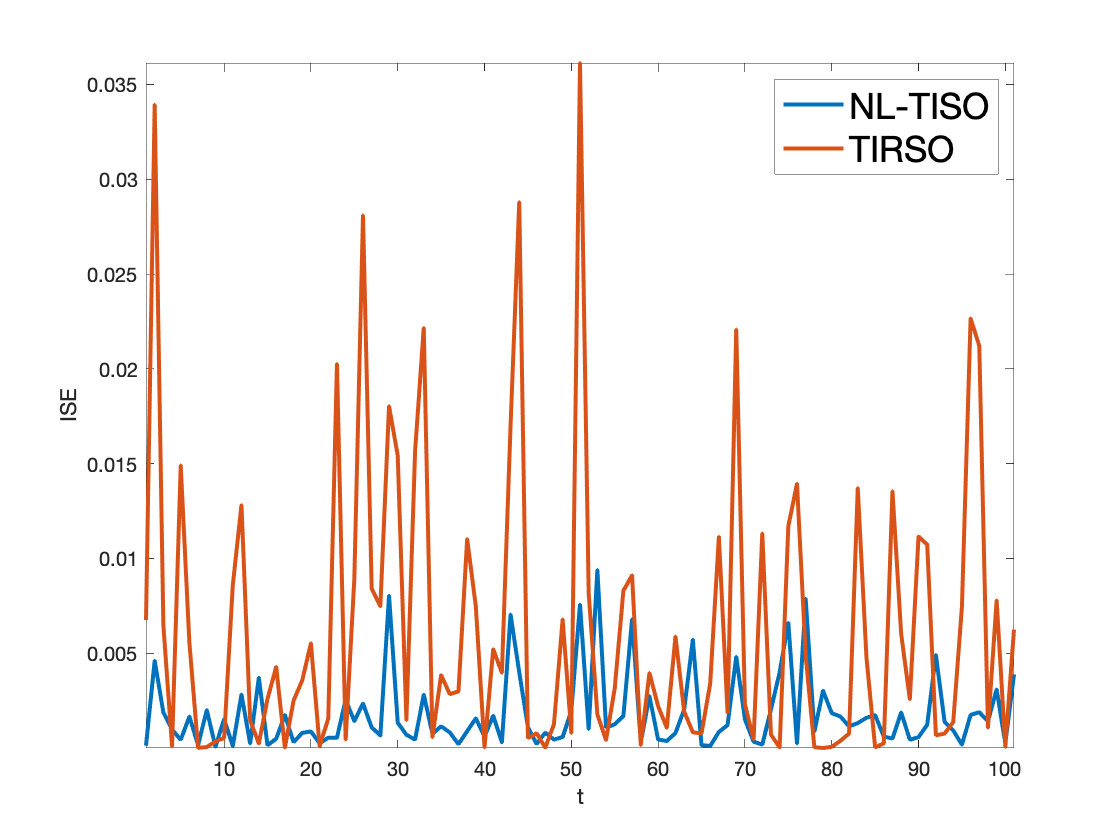}}
\caption{ISE comparison of NL-TISO and TIRSO when the signal to be reconstructed is rapidly  varying.}
\label{error}
\end{figure}
\cB{The causal dependencies $\cbr{{\bs \alpha}_{n,\n}^{(p)}[t]}$ are estimated using the proposed NL-TISO algorithm using Gaussian kernel having variance $0.1$ and with hyper-parameters  $\lambda=0.1$ and $\gamma=10$. Since a stationary topology is considered in this experiment, we compute the $\ell_2$ norms $\widehat{b}_{n,n'}^\p=\|\bs{\alpha}_{n,n'}^\p[t]\|_2$ at $t=T$ and arrange them in a matrix structure similar to the graph adjacency matrix to visualize the causal dependencies. A similar strategy is adopted for the FGD and the TIRSO algorithms, and the estimated adjacency matrix is used to visualize the dependencies. The true and the estimated dependencies are shown in \cref{topid}, in which for each subplot, the $5\times5$ dependency matrices corresponding to $p=1 \rm{~and~}2$ are concatenated, resulting in a size $10\times 5$ size matrix. From \cref{topid}, it is clear that the NL-TISO algorithm outperforms others in identifying the causal relationship. }
\subsubsection{Signal Reconstruction Experiment}
\label{srec}
\cB{In this experiment, using the inferred causal dependencies, we reconstruct the time series and compare it with the true signals. In contrary to the previous experiment, a dynamic graph-topology is considered here using a time varying adjacency matrix \quad
\begin{align}\label{eq:sine1}
    a^\p_{n,n'}[t+1]=a^\p_{n,n'}[t]+0.01\sin(0.03*t)
\end{align}
with a random initialization.
We use a different non-linear dependency compared to the previous experiment to generate data:
% \resizebox{1\linewidth}{!}{
% 	\begin{minipage}{1\linewidth}
\begin{align}
    f_{n,n'}^{(p)}(x)=0.4\sin(\pi x^2)+0.3\sin(2\pi x)+0.3\sin(3\pi x).\label{eq:sine2}
\end{align}
% \end{minipage}
% }
Graph-connected time series ($N=5$) are generated using \eqref{eq:var}, \eqref{eq:sine1}, and \eqref{eq:sine2} in a similar manner as described in \cref{Tid}.} 

\cB{The causal dependencies $\cbr{\boldsymbol{\alpha}_{n,n'}^\p}$ are estimated from the time series using NL-TISO with a Gaussian kernel having variance $0.02$ and with hyper-parameters $\lambda=10^{-6}$ and $\gamma=10$ . Using the same Gaussian kernel and the estimated dependencies, the time series are reconstructed. In \cref{rec}, a visual comparison of both the true and reconstructed time series at one of the five nodes is shown. We observed that the reconstructed signal is very close to the true one, although a Gaussian-based kernel is used to infer the non-linearity imposed by \eqref{eq:sine2}, which in turn indicates that kernel-based representations are a powerful tool in handling the non-linear causal dependencies. Further, the signal reconstruction quality of the state-of-the-art algorithms TIRSO \cite{onlzam2019} and FGD \cite{only.s2018} are compared using \textit{instantaneous squared error}, which is defined as    $ISE(t)=(y_n(t)-\hat{y}_n(t))^2$ and is plotted in \cref{error}, which concludes that NL-TISO outperforms TIRSO by a considerable margin for the non-linear signal models. We have also observed that the ISE of the FGD algorithm is much worse than NL-TISO and TIRSO and is not shown in the figure.}

\subsection{Experiments using Real Data}
\label{real}
\cB{In this section, we present experiments using real data collected from Lundin’s offshore oil and gas (O\&G) platform Edvard-Grieg\footnote{https://www.lundin-energy.com/}. We consider a directed graph with $24$ nodes; each node corresponds to  temperature (T), pressure (P), or oil-level (L) sensors. These sensors are placed in the separators of  decantation tank that separates oil, gas, and water. The time series are obtained by  uniformly sampling the sensor readings and applying normalization to have zero mean and unit sample variance. These time series are expected to exhibit causal dependencies due to the underlying physical coupling arising from the pipeline connections and the control systems.}

\begin{figure}[hbt!]
\centerline{\includegraphics[scale=0.17]{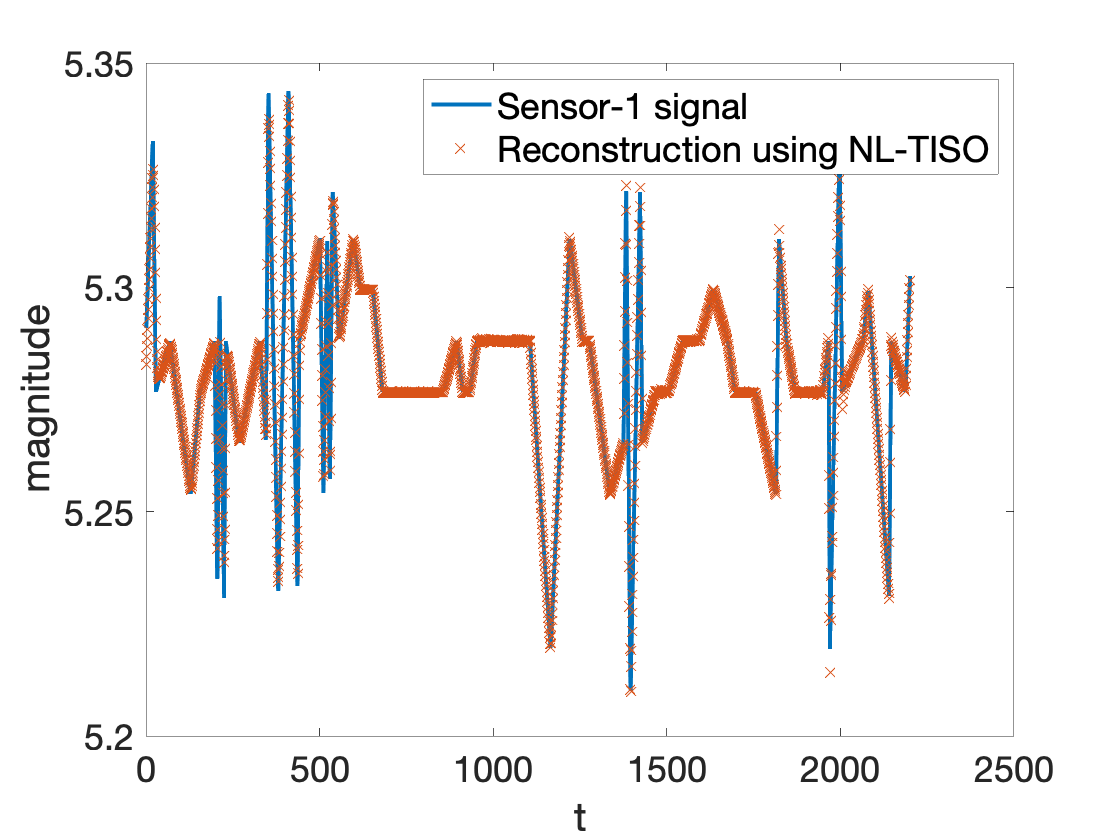}}
\caption{Reconstruction of sensor-1 signal with sampling time $5$s from Lundin data  using estimated causal dependencies.}
\label{realpred}
\end{figure}

\begin{figure}[hbt!]
\centerline{\includegraphics[scale=0.17]{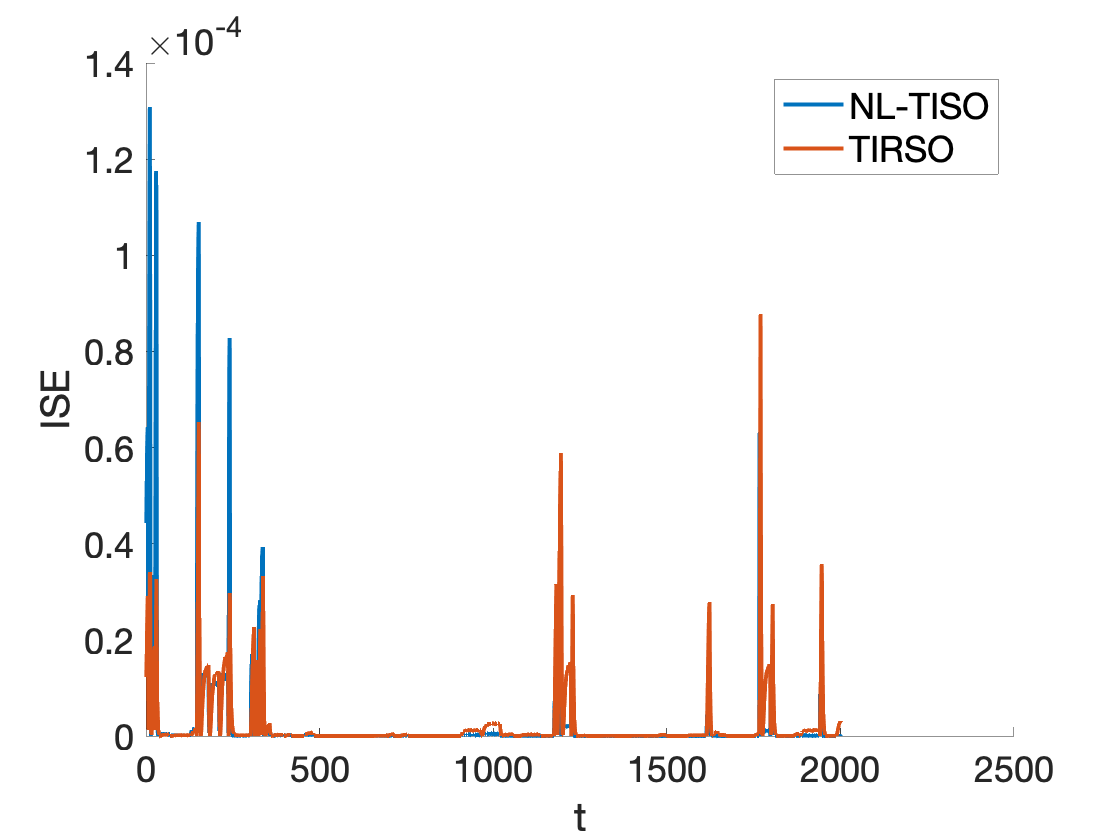}}
\caption{ISE comparison of NL-TISO and TIRSO using real data.}
\label{realerror}
\end{figure}

\begin{figure}[hbt!]
\centerline{\includegraphics[scale=0.17]{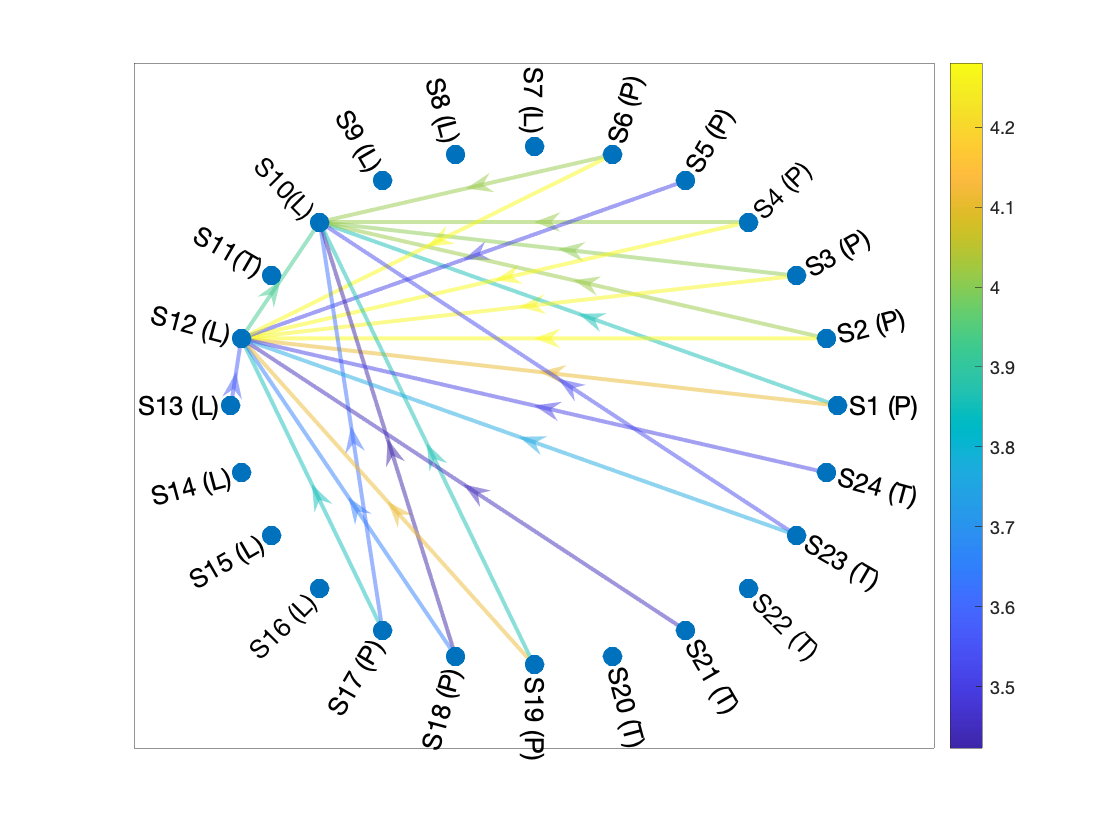}}
\caption{Causality graph in oil and gas plant estimated by NL-TISO}
\label{realtop}
\end{figure}

\cB{The causal dependencies are learned using NL-TISO with a Gaussian kernel having a variance of $0.1$ and with hyper parameter values $\lambda=0.1$ and $\gamma=10$. In \cref{realpred}, we show one portion of the reconstructed signal corresponding to  sensor-1, which is a pressure sensor, and it can be observed that the reconstructed signal is very close to the true sensor reading.} Further, in Fig.\ref{realerror}, we compare the reconstruction error of NL-TISO with TIRSO in terms of ISE for sensor-1 signal samples. We observe that NL-TISO outperforms TIRSO by a considerable margin, which supports the effectiveness of proposed algorithm in learning real world topology. The causal dependencies among the $24$ time series obtained by averaging the NL-TISO estimates for one hour is shown in \cref{realtop}. 


\section{Conclusion}
An online algorithm for non-linear topology identification from graph-connected time-series was proposed in this paper. Most of the state-of-the-art algorithms solve the topology estimation problem by assuming a linear and stationary topology. However, many real-world networks are highly dynamic and non-linear. The proposed algorithm, NL-TISO, is devised based on kernel representation to handle the non-linearities of the real-world sensor networks. Further, using a composite objective mirror descent method, NL-TISO estimates sparse topology in an online fashion aiming at dynamic system models. Qualitative and quantitative empirical evidence provided in the paper using real and synthetic data show that NL-TISO is an effective algorithm to infer the causal dependencies of real-world sensor networks. We identify two major limitations of the proposed framework: \textbf{i)} the computational complexity and memory requirements of kernel-based representations increases considerably with number of data points which is handled in NL-TISO by considering a time window to select recent samples and \textbf{ii)} the variance of the Gaussian kernels used in NL-TISO are heuristically chosen. These limitations could be handle by further research on dictionary-based multi-kernel representations, which will be devoted to our future work.

% \bibitem{b1} G. Eason, B. Noble, and I. N. Sneddon, ``On certain integrals of Lipschitz-Hankel type involving products of Bessel functions,'' Phil. Trans. Roy. Soc. London, vol. A247, pp. 529--551, April 1955.
% \bibitem{b2} J. Clerk Maxwell, A Treatise on Electricity and Magnetism, 3rd ed., vol. 2. Oxford: Clarendon, 1892, pp.68--73.
% \bibitem{b3} I. S. Jacobs and C. P. Bean, ``Fine particles, thin films and exchange anisotropy,'' in Magnetism, vol. III, G. T. Rado and H. Suhl, Eds. New York: Academic, 1963, pp. 271--350.
% \bibitem{b4} K. Elissa, ``Title of paper if known,'' unpublished.
% \bibitem{b5} R. Nicole, ``Title of paper with only first word capitalized,'' J. Name Stand. Abbrev., in press.
% \bibitem{b6} Y. Yorozu, M. Hirano, K. Oka, and Y. Tagawa, ``Electron spectroscopy studies on magneto-optical media and plastic substrate interface,'' IEEE Transl. J. Magn. Japan, vol. 2, pp. 740--741, August 1987 [Digests 9th Annual Conf. Magnetics Japan, p. 301, 1982].
% \bibitem{b7} M. Young, The Technical Writer's Handbook. Mill Valley, CA: University Science, 1989.
% \end{thebibliography}
\vspace{12pt}

\bibliographystyle{IEEEtran}
\bibliography{bibliography}

\begin{thebibliography}{10}
\providecommand{\url}[1]{#1}
\csname url@samestyle\endcsname
\providecommand{\newblock}{\relax}
\providecommand{\bibinfo}[2]{#2}
\providecommand{\BIBentrySTDinterwordspacing}{\spaceskip=0pt\relax}
\providecommand{\BIBentryALTinterwordstretchfactor}{4}
\providecommand{\BIBentryALTinterwordspacing}{\spaceskip=\fontdimen2\font plus
\BIBentryALTinterwordstretchfactor\fontdimen3\font minus
  \fontdimen4\font\relax}
\providecommand{\BIBforeignlanguage}[2]{{%
\expandafter\ifx\csname l@#1\endcsname\relax
\typeout{** WARNING: IEEEtran.bst: No hyphenation pattern has been}%
\typeout{** loaded for the language `#1'. Using the pattern for}%
\typeout{** the default language instead.}%
\else
\language=\csname l@#1\endcsname
\fi
#2}}
\providecommand{\BIBdecl}{\relax}
\BIBdecl

\bibitem{leax.d2019}
X.~{Dong}, D.~{Thanou}, M.~{Rabbat}, and P.~{Frossard}, ``Learning graphs from
  data: A signal representation perspective,'' \emph{IEEE Signal Processing
  Magazine}, vol.~36, no.~3, pp. 44--63, 2019.

\bibitem{onlB.Z2017}
B.~{Zaman}, L.~M. {Lopez-Ramos}, D.~{Romero}, and B.~{Beferull-Lozano},
  ``Online topology estimation for vector autoregressive processes in data
  networks,'' in \emph{2017 IEEE 7th International Workshop on Computational
  Advances in Multi-Sensor Adaptive Processing (CAMSAP)}, 2017, pp. 1--5.

\bibitem{dynl.m2018}
L.~{Lopez-Ramos}, D.~{Romero}, B.~{Zaman}, and B.~{Beferull-Lozano}, ``Dynamic
  network identification from non-stationary vector autoregressive time
  series,'' in \emph{2018 IEEE Global Conference on Signal and Information
  Processing (GlobalSIP)}, Nov 2018, pp. 773--777.

\bibitem{pata.c2018}
A.~{Chatterjee}, R.~J. {Shah}, and S.~{Sen}, ``Pattern matching based
  algorithms for graph compression,'' in \emph{2018 Fourth International
  Conference on Research in Computational Intelligence and Communication
  Networks (ICRCICN)}, 2018, pp. 93--97.

\bibitem{equc.j2011}
C.~J. {Quinn}, N.~{Kiyavash}, and T.~P. {Coleman}, ``Equivalence between
  minimal generative model graphs and directed information graphs,'' in
  \emph{2011 IEEE International Symposium on Information Theory Proceedings},
  2011, pp. 293--297.

\bibitem{topg.b2018}
G.~Giannakis, Y.~Shen, and G.~Karanikolas, ``Topology identification and
  learning over graphs: Accounting for nonlinearities and dynamics,''
  \emph{Proceedings of the IEEE}, vol. 106, no.~5, pp. 787--807, May 2018.

\bibitem{nonshe2019}
Y.~Shen, G.~B. Giannakis, and B.~Baingana, ``Nonlinear structural vector
  autoregressive models with application to directed brain networks,''
  \emph{IEEE Transactions on Signal Processing}, vol.~67, pp. 5325--5339, 2019.

\bibitem{onlzam2019}
B.~{Zaman}, L.~M.~L. {Ramos}, D.~{Romero}, and B.~{Beferull-Lozano}, ``Online
  topology identification from vector autoregressive time series,'' \emph{IEEE
  Transactions on Signal Processing}, vol.~69, pp. 210--225, 2021.

\bibitem{aspnoa2013}
N.~Simon, J.~Friedman, T.~Hastie, and R.~Tibshirani, ``A sparse-group lasso,''
  \emph{Journal of Computational and Graphical Statistics}, vol.~22, no.~2, pp.
  231--245, 2013.

\bibitem{vecche2011}
\BIBentryALTinterwordspacing
G.~Chen, D.~R. Glen, Z.~S. Saad, J.~Paul~Hamilton, M.~E. Thomason, I.~H.
  Gotlib, and R.~W. Cox, ``Vector autoregression, structural equation modeling,
  and their synthesis in neuroimaging data analysis,'' \emph{Computers in
  biology and medicine}, vol.~41, no.~12, p. 1142—1155, December 2011.
  [Online]. Available: \url{https://europepmc.org/articles/PMC3223325}
\BIBentrySTDinterwordspacing

\bibitem{unimic2006}
C.~A. Micchelli, Y.~Xu, and H.~Zhang, ``Universal kernels,'' \emph{J. Mach.
  Learn. Res.}, vol.~7, p. 2651–2667, Dec. 2006.

\bibitem{neuale2018}
A.~Tank, I.~Covert, N.~Foti, A.~Shojaie, and E.~Fox, ``Neural granger causality
  for nonlinear time series,'' \emph{arXiv:1802.05842}, 2018.

\bibitem{kery.s2018}
Y.~{Shen}, B.~{Baingana}, and G.~B. {Giannakis}, ``Kernel-based structural
  equation models for topology identification of directed networks,''
  \emph{IEEE Transactions on Signal Processing}, vol.~65, no.~10, pp.
  2503--2516, 2017.

\bibitem{leasch2001}
B.~Scholkopf and A.~J. Smola, \emph{Learning with Kernels: Support Vector
  Machines, Regularization, Optimization, and Beyond}.\hskip 1em plus 0.5em
  minus 0.4em\relax Cambridge, MA, USA: MIT Press, 2001.

\bibitem{only.s2018}
Y.~{Shen} and G.~B. {Giannakis}, ``Online identification of directional graph
  topologies capturing dynamic and nonlinear dependencies†,'' in \emph{2018
  IEEE Data Science Workshop (DSW)}, 2018, pp. 195--199.

\bibitem{onlm.m2020}
M.~Moscu, R.~Borsoi, and C.~Richard, ``Online kernel-based graph topology
  identification with partial-derivative-imposed sparsity,'' in \emph{Signal
  Processing (EUSIPCO), 28th European Conference on}, 2020.

\bibitem{Comjoh2010}
J.~C. Duchi, S.~Shalev-shwartz, Y.~Singer, and A.~Tewari, ``Composite objective
  mirror descent,'' in \emph{In Adam Tauman Kalai and Mehryar Mohri, editors,
  Proc. of the 23th Annual Conference on Learning Theory (COLT’10}, 2010, pp.
  14--26.

\bibitem{Wahba1990}
G.~Wahba, \emph{Spline Models for Observational Data}.\hskip 1em plus 0.5em
  minus 0.4em\relax SIAM Press, Society for Industrial and Applied Mathematics,
  1990.

\bibitem{ageolk2000}
B.~Olkopf, R.~Herbrich, A.~Smola, and R.~Williamson, ``A generalized
  representer theorem,'' \emph{Computational Learning Theory}, vol.~42, 06
  2000.

\bibitem{bregut2011}
M.~U. Gutmann and J.-i. Hirayama, ``Bregman divergence as general framework to
  estimate unnormalized statistical models,'' in \emph{Proceedings of the
  Twenty-Seventh Conference on Uncertainty in Artificial Intelligence}, ser.
  UAI'11.\hskip 1em plus 0.5em minus 0.4em\relax Arlington, Virginia, USA: AUAI
  Press, 2011, p. 283–290.

\bibitem{amupui2009}
A.~Puig, A.~Wiesel, and A.~Hero, ``A multidimensional shrinkage-thresholding
  operator,'' vol.~18, 10 2009, pp. 113 -- 116.

\end{thebibliography}
\end{document}